\documentclass[sigconf]{acmart}
\AtBeginDocument{%
  }

\usepackage{algorithm}
\usepackage{algorithmic}

\copyrightyear{2026}
\acmYear{2026}
\setcopyright{cc}
\setcctype{by}
\acmConference[SIGIR '26]{Proceedings of the 49th International ACM SIGIR Conference on Research and Development in Information Retrieval}{July 20--24, 2026}{Melbourne, VIC, Australia}
\acmBooktitle{Proceedings of the 49th International ACM SIGIR Conference on Research and Development in Information Retrieval (SIGIR '26), July 20--24, 2026, Melbourne, VIC, Australia}
\acmDOI{10.1145/3805712.3808395}
\acmISBN{979-8-4007-2599-9/2026/07}

\begin{document}
\title{When \& How to Write for Personalized Demand-aware Query Rewriting in Video Search}

\author{Cheng Cheng}
\authornote{Both authors contributed equally to this research.}
\email{trcheng@tencent.com}
\author{Chenxing Wang}
\authornotemark[1]
\authornote{Corresponding author.}
\email{magnuswang@tencent.com}
\affiliation{%
  \institution{Weixin Group, Tencent}
  \country{China}
}

\author{Aolin Li}
\email{churenli@tencent.com}
\author{Haijun Wu}
\email{ethanhjwu@tencent.com}
\affiliation{%
  \institution{Weixin Group, Tencent}
  \country{China}
  }

\author{Huiyun Hu}
\authornotemark[1]
\email{huiyunhu@tencent.com}
\author{Juyuan Wang}
\email{jovwang@tencent.com}
\author{Dongliang Liao}
\email{liaodl258@gmail.com}
\affiliation{%
  \institution{Weixin Group, Tencent}
  \country{China}
  }

\renewcommand{\shortauthors}{Cheng et al.}

\begin{abstract}
In video search systems, user historical behaviors provide rich context for identifying search intent and resolving ambiguity. 
However, traditional methods utilizing implicit history features often suffer from signal dilution and delayed feedback. 
To address these challenges, we propose WeWrite, a novel Personalized Demand-aware Query Rewriting framework. 
Specifically, WeWrite tackles three key challenges: 
(1) When to Write: An automated posterior-based mining strategy extracts high-quality samples from user logs, identifying scenarios where personalization is strictly necessary;
(2) How to Write: A hybrid training paradigm combines Supervised Fine-Tuning (SFT) with Group Relative Policy Optimization (GRPO) to align the LLM's output style with the retrieval system;
(3) Deployment: A parallel "Lightweight Recall" architecture ensures low latency.
Extensive offline experiments demonstrate that WeWrite significantly outperforms general-purpose LLMs (e.g., Qwen3-32B), improving rewriting accuracy by over 27\%.
We deployed WeWrite in the main scene of WeChat, a large-scale mobile platform with over 1 billion monthly active users, which improves the Click-Through Video Volume (VV$>$10s) by 1.07\% and reduces the Query Reformulation Rate by 2.97\% without increasing the serving cost.
\end{abstract}

\ccsdesc[500]{Information systems~Novelty in information retrieval}

\keywords{Query Rewriting, Large Language Models, Video Search}


\maketitle

\section{Introduction}
\label{sec:intro}

\begin{figure*}[t]
  \centering
  \includegraphics[width=\linewidth]{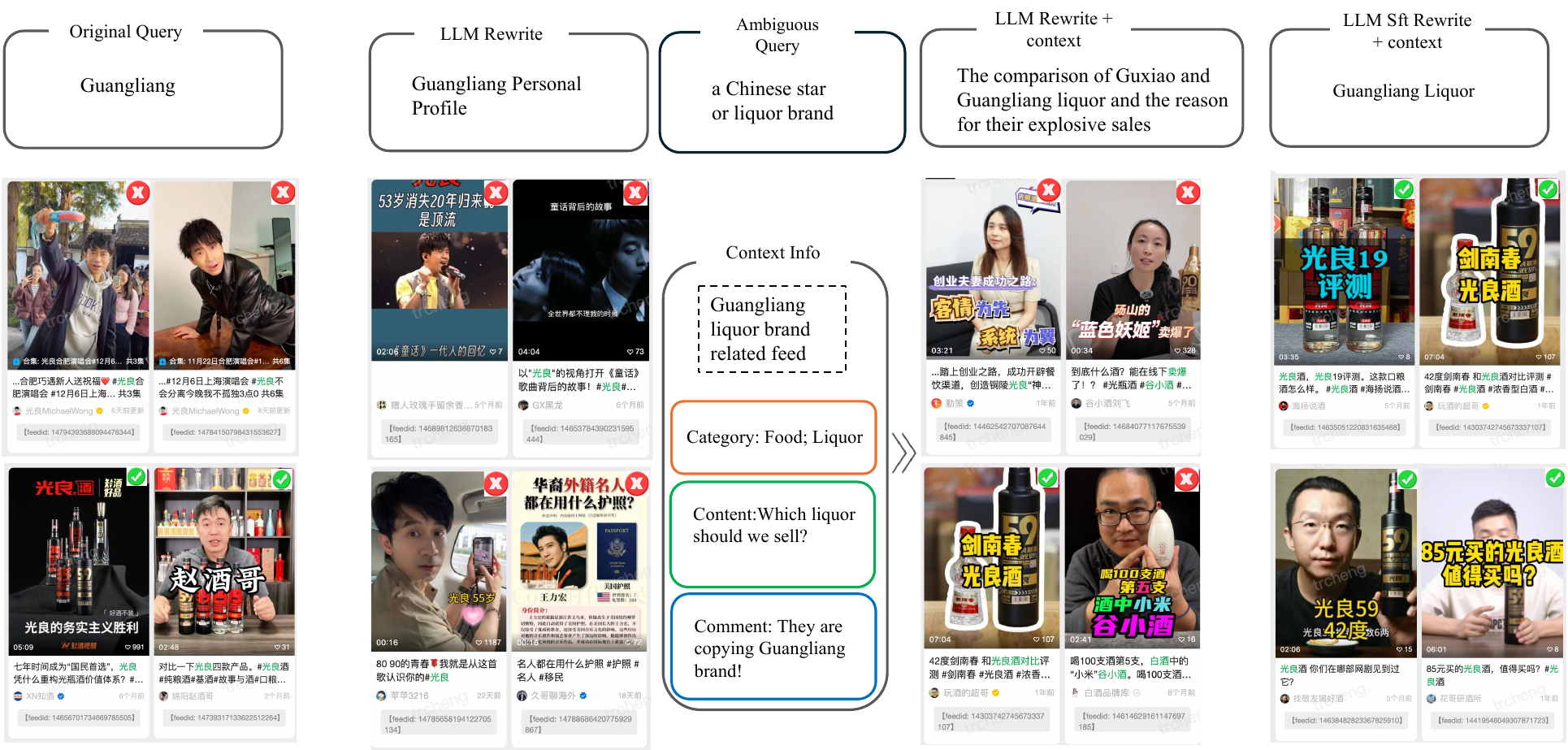} 
  \caption{Positive Case: WeWrite resolves ambiguity (Singer vs. Liquor) using user history.}
  \label{fig:motivating_example}
\end{figure*}

Short-form video platforms (e.g., WeChat\footnote{The world's largest standalone mobile app with over 1 billion monthly active users.} Channels) rely heavily on search engines for content discovery. 
However, user queries are often brief and ambiguous. 
As shown in Figure~\ref{fig:motivating_example}, the query "Guang Liang" can refer to a singer or a liquor brand. 
Generic search engines fail to satisfy users with specific latent interests (e.g., liquor), highlighting the necessity of Personalized Query Rewriting.

\begin{figure}[t]
  \centering
  \includegraphics[width=\linewidth]{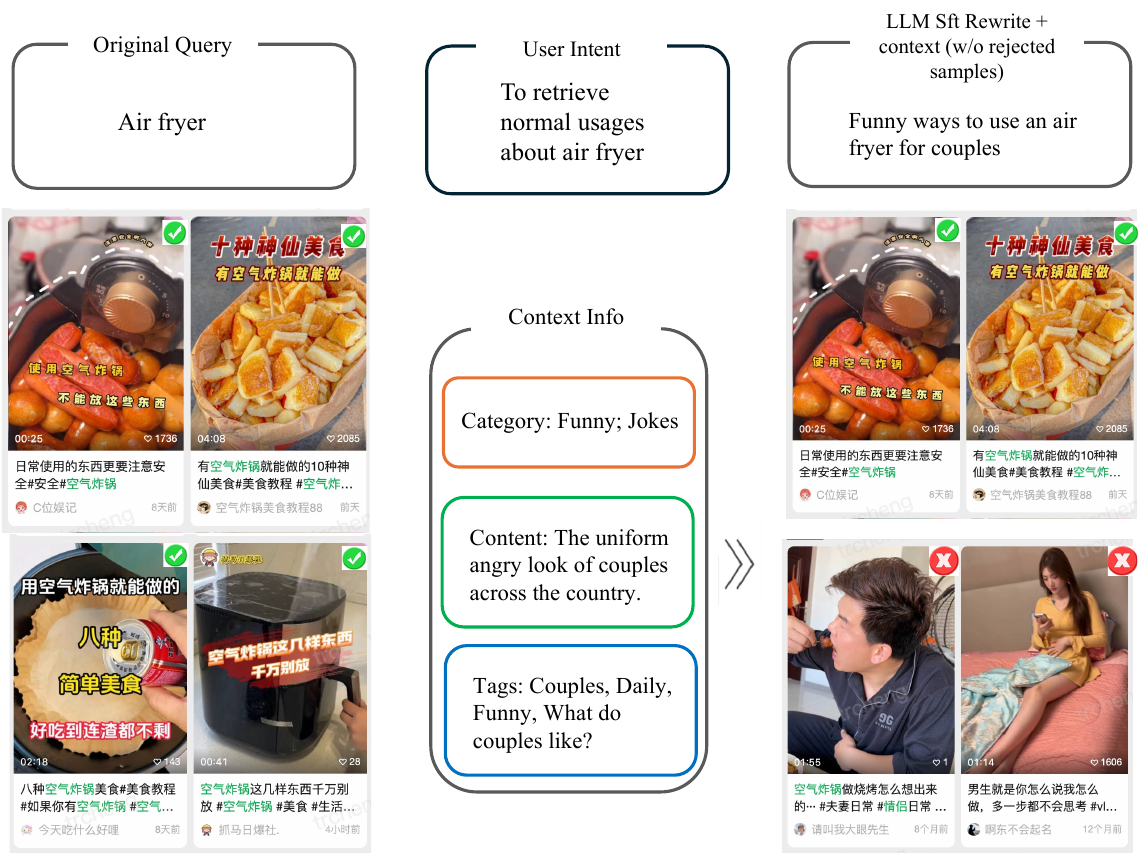} 
  \caption{Negative Case: Indiscriminate rewriting causes \textbf{intent drift}. Functional queries ("Air fryer") should not be rewritten based on entertainment history.}
  \label{fig:intent_drift}
  \vspace{-20pt}
\end{figure}


Existing rewriting approaches have evolved from discriminative paradigms~\cite{li2022query,xiao2019weakly,wang2023learning} to generative paradigms. Within the generative landscape, while some methods leverage LLMs to synthesize relevant external knowledge for query expansion~\cite{wang2023query2doc, baek2025crafting, jagerman2023query}, industrial systems increasingly focus on refining queries to better align with retrieval objectives.  
Some frameworks include CardRewriter~\cite{gong2025cardrewriter} for long-tail queries, alongside specialized industrial frameworks like Taobao's semantic alignment~\cite{peng2024large, feng2025complicated}, Alipay's adaptive multi-agent system~\cite{zheng2025maaqr}, and Meituan's IterQR~\cite{chen2025iterqr}.
To further bridge the gap between text generation and retrieval effectiveness, recent studies have integrated Reinforcement Learning (RL) algorithms—including PPO~\cite{schulman2017proximal}, DPO~\cite{rafailov2023direct}, KTO~\cite{ethayarajh2024kto}, GRPO~\cite{shao2024deepseekmath},among others—to align LLM outputs with downstream system rewards~\cite{dai2024enhancing, zhang2025value, nguyen2025minielm}.
However, despite these advances, these methods remain predominantly \textit{context-agnostic}, treating all users identically and overlooking idiosyncratic search intents. Although personalization has been extensively explored in related fields like Query Auto-Completion (QAC)~\cite{bar2011context, yin2020learning, maurya2023trie, wang2025personalized} and Query Suggestion~\cite{guo2025onesug,baek2024knowledge,zhou2022personalized}, applying explicit personalized rewriting to real-time search presents unique, non-trivial challenges: 
(1) \textbf{When \& How to Write:} Indiscriminate rewriting risks intent drift (see Figure~\ref{fig:intent_drift}), where functional queries are wrongly biased by historical interests. The core challenge is identifying \textit{when} personalization is strictly necessary and \textit{how} to generate rewrites that align with the specific semantic space of the retrieval system's index. 
(2) \textbf{System Latency:} The high inference cost of LLMs prohibits direct integration into the synchronous path of real-time video search systems with strict latency constraints.

To this end, we propose WeWrite, a novel Personalized Demand-aware Query Rewriting framework. Our primary contributions are summarized as follows: (1) Posterior-based "When" Strategy: We introduce an automated mining mechanism that utilizes user posterior feedback to identify scenarios where personalization is strictly necessary, effectively mitigating intent drift.
(2) style-aligned "How" Training: We propose a hybrid training paradigm combining SFT with GRPO. By optimizing for retrieval-oriented rewards (Index Hit Rate and CTR), we align the LLM's output style with the system index.
(3) Lightweight Recall Deployment: We devise a parallel architecture that decouples LLM inference from the main path, ensuring low latency.
Offline experiments show that WeWrite improves rewriting accuracy by over 27\% compared to general-purpose LLMs, while online A/B testing on a large-scale video platform demonstrates a 1.07\% increase in Click-Through Video Volume (VV$>$10s) and a 2.97\% reduction in Query Reformulation Rate.

\section{Methodology}
\label{sec:method}

\begin{figure*}[t]
  \centering
\includegraphics[width=0.75\textwidth]{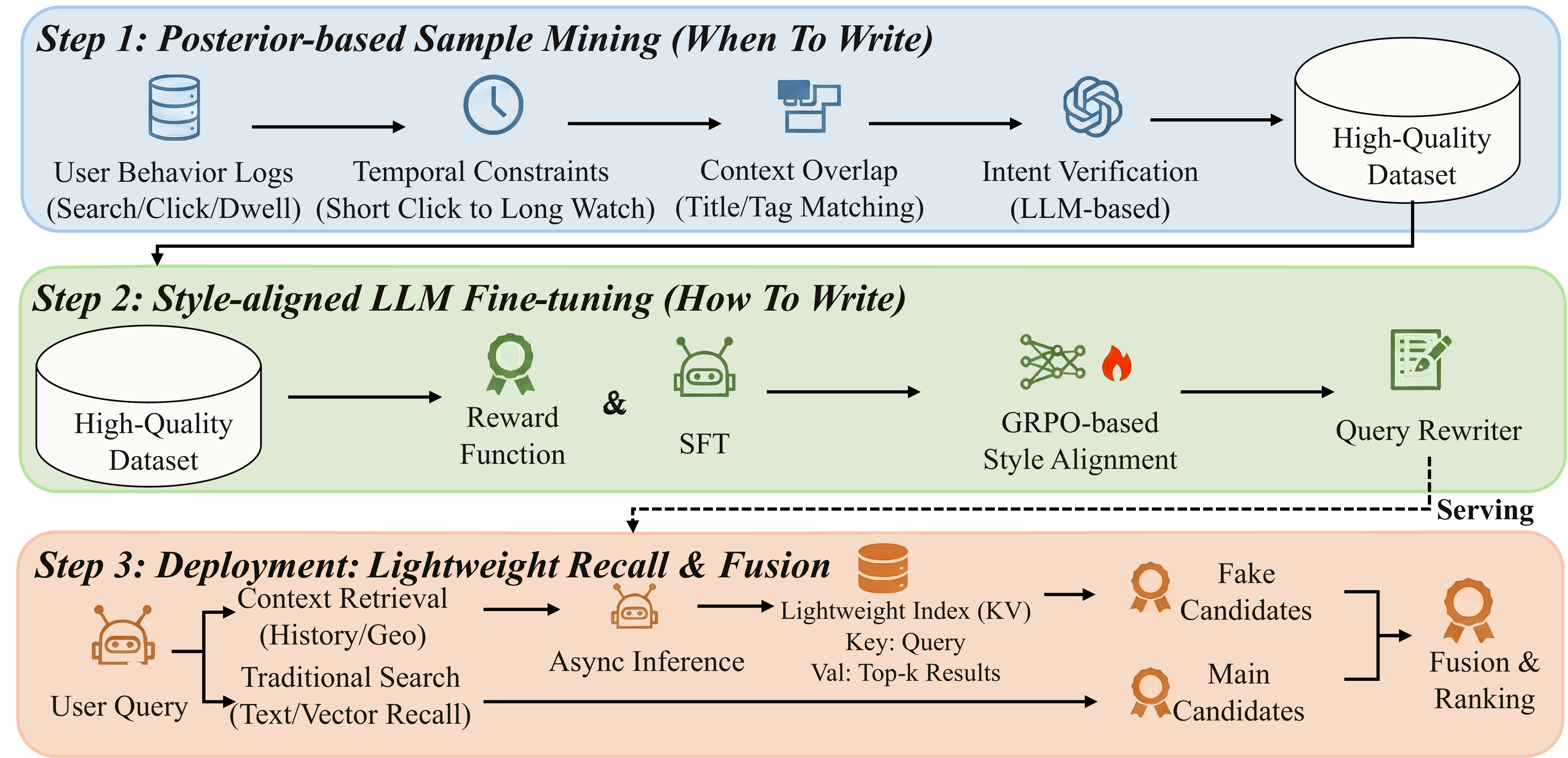}  \caption{Overview of the proposed framework. It comprises offline mining of intent-aligned samples, hybrid LLM training (SFT+RL) for style alignment, and an online parallel "Lightweight Recall" architecture to minimize latency.}
  \label{fig:framework}
\end{figure*}

In this section, we present the proposed Personalized Demand-aware Query Rewriting framework. 
As illustrated in Figure~\ref{fig:framework}, the system consists of three main modules: 
(1) Posterior-based Sample Mining, which constructs high-quality training data from user feedback logs to determine when to write; 
(2) Style-aligned LLM Fine-tuning, which trains the model to learn how to write by combining SFT and GRPO; 
(3) Deployment with Lightweight Recall, a parallel execution architecture ensuring low-latency online serving.

\subsection{Problem Formulation}
The goal of personalized query rewriting is to generate a reformulated query $Q'$ based on the user's original query $Q$ and historical context $C_u$, such that $Q'$ retrieves documents maximizing the user's utility (e.g., valid video consumption).
Formally, we define the user context as a tuple $C_u = \{ H_{query}, H_{video}, G \}$, where $H_{query}$ and $H_{video}$ denote the sequence of historical queries and watched videos, respectively, and $G$ represents geolocation features.

We formulate the rewriting task as learning a conditional generation policy $\pi_\theta(Q' | Q, C_u)$. 
Unlike traditional rewriting that focuses solely on semantic equivalence, our objective is to maximize the expected reward reflecting both \textit{user satisfaction} (Click-Through Rate) and \textit{system retrievability} (Index Hit Rate):
\begin{equation}
    \theta^* = \operatorname*{argmax}_\theta \mathbb{E}_{(Q, C_u) \sim \mathcal{D}, Q' \sim \pi_\theta} [ R(Q', \mathcal{I}_{sys}) ]
\end{equation}
where $\mathcal{I}_{sys}$ represents the online search index and $R(\cdot)$ is the reward function defined in Section~\ref{sec:model}.
\subsection{Posterior-based Sample Mining}
\label{sec:mining}
A core challenge in personalized rewriting is determining "When to Write". 
Indiscriminately rewriting every query introduces noise and increases system load. 
To address this, we propose an automated mining strategy that leverages user posterior behaviors to extract samples where personalization is strictly necessary, yielding rewrite (positive) and reject (negative) data.
\subsubsection{Positive Sample Mining (Rewrite)}
User dissatisfaction is typically signaled by rapid query reformulations. We capture the transition from a failed query $Q_{orig}$ to a successful reformulation $Q_{next}$, conditioned on $C_u$. A pair $(Q_{orig}, Q_{next})$ is a candidate if it meets two temporal constraints: 
(1) \textbf{Origin Failure}: Minimal interaction with $Q_{orig}$ results ($T_{orig} < 2.4s$). 
(2) \textbf{Next Success}: Valid consumption on $Q_{next}$ results ($T_{next} > 10s$).

To exclude non-personalized reformulations, we apply a two-stage demand-aware filter. \textbf{Stage 1: Context Overlap.} We retain samples only if gain words $\Delta T = \text{Terms}(Q_{next}) \setminus \text{Terms}(Q_{orig})$ intersect with user history $C_u$ (e.g., $H_{video}$ titles), ensuring intent shifts are grounded in behavior. \textbf{Stage 2: LLM Verification.} To resolve subtle dependencies, Qwen3-32B acts as an expert annotator to verify if the reformulation is explicitly supported by $C_u$ via a binary classification prompt:
\begin{quote}
\small
\textit{\textbf{Role:}} You are an experienced data annotator specializing in short-form video search analysis. \\
\textit{\textbf{Task:}} Users often reformulate queries when dissatisfied. Given the \texttt{Original Query}, \texttt{User Context} (Gender, Location, Viewed Videos), and the \texttt{Reformulated Query}, determine if the reformulation is derived from the context or is a valid correction. \\
\textit{\textbf{Constraint:}} A valid contextual dependency exists \textbf{only if} the new terms (gain words) in the \texttt{Reformulated Query} explicitly appear in the \texttt{User Context}. \\
\textit{\textbf{Input:}}
\begin{itemize}
    \item Original Query: $Q_{orig}$
    \item User Context: $C_u$
    \item Reformulated Query: $Q_{next}$
\end{itemize}
\textit{\textbf{Output:}} Answer "Positive" if there is a clear contextual clue; otherwise, answer "Negative".
\vspace{-5pt}
\end{quote}
Only "Positive" labeled samples form the final training set $S_{pos} = \{(C_u, Q_{orig}) \to Q_{next}\}$.
\subsubsection{Negative Sample Mining (Reject)}
To teach the model "when not to rewrite," we mine $S_{neg} = \{(C_u, Q_{orig}) \to \texttt{<reject>}\}$ from two sources:
(1) \textbf{Direct Satisfaction}: Cases where $Q_{orig}$ yielded long consumption ($T_{orig} > \tau_{long}$) without subsequent reformulations, indicating the original intent was fulfilled. 
(2) \textbf{Personalization Redundancy}: Instances rejected by the aforementioned context-overlap or LLM-based filters, indicating that a rewrite lacks sufficient contextual grounding. 
The model is thus trained to output a special \texttt{<reject>} token, preserving the original query and maintaining system efficiency.

\subsection{Style-aligned LLM Fine-tuning}
\label{sec:model}

After determining "when" to rewrite, the next challenge is "how" to generate rewrites that are both semantically accurate and system-friendly. 
We employ a two-stage training paradigm: SFT for semantic generation and GRPO for style alignment.
Specifically, we use the mined dataset $D_{train} = S_{pos} \cup S_{neg}$ to fine-tune a base LLM (e.g., Qwen3 \cite{yang2025qwen3}). 
To guide the model's generation, we construct an input sequence $x$ that incorporates the user context $C_u$ and the original query $Q_{orig}$ wrapped in a specific system instruction. 
The prompt is designed as follows:

\begin{quote}
\small
\textit{\textbf{Instruction:}} Based on user features such as recently watched videos, gender, and historical search queries, determine whether current query needs to be refined, rewritten, or corrected. If necessary, directly output the rewritten or corrected query; otherwise, output \texttt{<reject>}.
\end{quote}

Correspondingly, the target sequence $y$ is either the reformulated query $Q_{next}$ (for positive samples) or the special token \texttt{<reject>} (for negative samples).
The SFT objective is to minimize the standard negative log-likelihood:
\begin{equation}
    \mathcal{L}_{SFT} = - \sum_{t=1}^{|y|} \log P(y_t | y_{<t}, C_u, Q_{orig})
\end{equation}
\subsubsection{GRPO-based Style Alignment}

A common issue with generative rewriting is that the model may produce semantically correct queries that are sufficiently rare or complex that they fail to retrieve valid results in the existing search index (e.g., "zero-recall" problem). 
To address this, we introduce a GRPO stage to align the generated queries with the online system's query distribution.

We model the rewriting process as a Markov Decision Process (MDP). 
The action space is the vocabulary of the LLM, and the state includes the context and generated tokens so far. 
To encourage the model to generate queries that are both semantically accurate and highly searchable, we design a hybrid posterior reward function $R(Q_{rew})$. 
Considering computational efficiency, we utilize ROUGE-L \cite{li2024unigen} as the evaluation score ($S_{eval}$) to measure the semantic similarity between the generated query $Q_{rew}$ and the ground-truth target $Q_{next}$.
The reward is formulated as:
\begin{equation}
    \label{eq:reward-design}
    R(Q_{rew}) = 
    \begin{cases} 
    1.0 + \alpha \cdot \text{clip}(\log(\text{qv} + 1), N_{low}, N_{up}) & \text{if } S_{eval} > \tau_{thd} \\
    S_{eval} & \text{otherwise}
    \end{cases}
\end{equation}
where $\text{qv}$ represents the historical query volume (search frequency) of $Q_{rew}$ in the system logs, serving as a proxy for indexability.
$\tau_{thd}$ is a threshold that ensures that the model prioritizes semantic fidelity before optimizing for search popularity.
$N_{low}$ and $N_{up}$ are clipping bounds used to stabilize the reward scale, and $\alpha$ is a scaling coefficient.
This mechanism guides model to generate popular, well-indexed queries \textit{only when} they accurately reflect user's intent.

Finally, we employ GRPO to optimize the policy without the overhead of a value network. 
For each input, we sample a group of $G$ rewrites $\{Q'_{1}, \dots, Q'_{G}\}$ and compute the advantage $A_i$ by normalizing the reward $R(Q'_i)$ against group statistics:
\begin{equation}
    A_i = \frac{R(Q'_i) - \mu(\mathbf{R})}{\sigma(\mathbf{R}) + \epsilon}
\end{equation}

The final training objective $\mathcal{L}_{Total}$ maximizes the policy advantage via a clipped surrogate loss $\mathcal{J}_{GRPO}(\theta)$ while maintaining stability through SFT regularization:
\begin{equation}
\mathcal{L}_{Total} = \beta \cdot \mathcal{L}_{SFT} - \mathbb{E} \left[ \mathcal{J}_{GRPO}(\theta) \right]
\end{equation}
{\small \begin{equation}
\mathcal{J}_{GRPO}(\theta) = \frac{1}{G} \sum_{i=1}^G \frac{1}{|o_{i}|} \sum_{t=1}^{|o_{i}|} \left( \min \left\{ r_{i,t}{A}_{i,t}, \mathrm{clip}(r_{i,t}, 1-\epsilon, 1+\epsilon){A}_{i,t} \right\} - \gamma D_{KL} \right)
\end{equation}}
where $r_{i,t}$ is the policy ratio, $\gamma$ is the KL penalty coefficient, and $\beta$ controls the contribution of the SFT loss.

\subsection{Deployment: Lightweight Recall \& Fusion}
\label{sec:deployment}

Deploying LLMs in a real-time video search system with strict latency constraints is non-trivial. 
A serial execution flow would introduce unacceptable delays. 
To solve this, we propose a parallel Lightweight Recall architecture.

\subsubsection{Lightweight Index Construction}
We pre-build a key-value index, denoted as $I_{lw}$, mapping valid system queries to top-performing documents. To ensure online availability, we perform daily update to the index (along with real-time update within each day).  To this end, the index can cover approximately 83\% incoming queries. Moreover, to mitigate the sparsity of click signals (especially for long-tail queries), we employ a hybrid construction strategy. First, for head queries with rich feedback, we cache the Top-$K$ documents based on historical CTR and dwell time. Second, for queries with sparse interaction data, we supplement the index by mining the top-ranked documents from historical search logs.
This pre-built index acts as a massive "result cache" ($K=50$), allowing system to retrieve candidates for LLM-generated query $Q_{rew}$ in $O(1)$ time, bypassing the heavy online retrieval chain.

\subsubsection{Online Parallel Fusion}
The online serving flow minimizes latency through parallelism. 
First, when a user request arrives, the system triggers the Traditional Search Path and the Personalized Rewriting Path simultaneously. The LLM generates the rewrite $Q_{rew}$ asynchronously. If $Q_{rew}$ hits the Lightweight Index $I_{lw}$, we retrieve the candidate document list $Cand_{lw}$. Moreover, to ensure safety, we apply a lightweight relevance model to filter $Cand_{lw}$, removing documents that are irrelevant to the original intent. Finally, $Cand_{lw}$ is merged with the main candidate set $Cand_{main}$.
By parallelizing LLM inference with the traditional recall phase, total end-to-end latency remains unaffected, achieving zero-perceived-latency personalization.

\section{Experiments}
\label{sec:exp}

\subsection{Offline Performance}

\subsubsection{Evaluation Setup and Metric}
For offline experiments, we utilize 20 consecutive days of online user logs as the training set, with the following day's data serving as the test set. To evaluate the similarity between generated rewrites and the ground truth, we employ BLEU-4~\cite{kim2025gure}, ROUGE-L~\cite{li2024unigen}, and Bert-Score~\cite{tanjim2025detecting} to evaluate lexical and semantic similarity. Furthermore, we define Acc as the proportion of samples where the weighted aggregate score exceeds 0.5:
\begin{equation}
    \text{Acc} = \frac{1}{N} \sum_{i=1}^{N} \mathbb{I} \left( \left( 0.4 \cdot B_i + 0.4 \cdot R_i + 0.2 \cdot S_i \right) > 0.5 \right)
\end{equation}
where $N$ denotes the total number of samples in the test set, $B_i, R_i, S_i$ are the respective scores for sample $i$, and $\mathbb{I}(\cdot)$ is the indicator function. The weights are empirically set to balance lexical and semantic contributions.

\subsubsection{Main Results and Ablation Analysis}

Table~\ref{tab:combined_results} summarizes the overall performance and ablation results. Specifically, for the main results, our proposed WeWrite significantly outperforms all baselines. Notably, task-specific adaptation on a 4B model yields a substantial performance leap over the 32B zero-shot/few-shot counterparts, improving \textit{Acc} by over 27\%. This confirms that LLM's emergent abilities alone are insufficient for specialized rewriting without domain adaptation. For the ablation on sample mining, incorporating LLM-based intent verification on top of the rule-based filtering process (comparing WeWrite (SFT) w/o Intent Verif. vs. SFT) boosts Acc from 0.500 to 0.535. This proves that semantic reasoning in posterior mining captures higher-quality training signals than rigid heuristics. For the ablation on reward design, the transition from \textit{GRPO-w/o qv} to our full model shows that incorporating popularity-based conditional rewards further improves \textit{Acc} to 0.583. Furthermore, we measure the query volume ($qv$) generated by the RL model relative to the SFT baseline. Experimental results show that the proportion of queries with improved $qv$ reaches 20.21\% when incorporating the $qv$ reward. This enhancement is primarily driven by two factors: (1) $qv$ serves as a reliable proxy for query quality, and (2) as an authentic metric from the production system, a higher $qv$ suggests that the generated queries better align with historical search distributions, thereby significantly strengthening the model's posterior capability. This indicates that the RL stage successfully aligns the model with real-world user search behaviors while maintaining semantic precision. 
\begin{table}[htbp]
    \centering
    \caption{Table 1: Performance comparison and ablation study. The top group compares WeWrite with baselines. The bottom group ablates LLM-based intent verification (via SFT variants) and the hybrid posterior reward design (via GRPO variants).}
    \label{tab:combined_results}
    \resizebox{\linewidth}{!}{
        \begin{tabular}{lccccc}
            \toprule
            Method & Acc & BLEU-4 & ROUGE-L & Bert-Score  \\
            \midrule
            Zero-shot (Qwen3-32B) w/o context & 0.111 & 0.078 & 0.196 & 0.616 \\
            Zero-shot (Qwen3-32B)             & 0.178 & 0.102 & 0.200 & 0.604 \\
            Few-shot (Qwen3-32B)              & 0.305 & 0.168 & 0.392 & 0.714 \\
            BART-Large (Fine-tuned)           & 0.298 & 0.163 & 0.388 & 0.708 \\
            \textbf{WeWrite}       & \textbf{0.583} & \textbf{0.330} & \textbf{0.620} & 0.828 \\
            \midrule
            \midrule
            WeWrite (SFT) w/o Intent Verif.   & 0.500 & 0.293 & 0.556 & 0.801 \\
            WeWrite (SFT)                     & 0.535 & 0.316 & 0.584 & 0.813 \\
            WeWrite (SFT-GRPO-w/o qv)      & 0.564 & 0.329 & 0.613 & \textbf{0.829} \\
            \bottomrule
        \end{tabular}
    }
    \vspace{-10pt}
\end{table}

\subsection{Online A/B Testing}

As shown in Table \ref{tab:ab-testing}, we deployed the optimal model (WeWrite) online. 
The A/B test results show a \textbf{1.07\% statistically significant increase (p<0.01)} in \textbf{click through video volumn with stay time $>$10s (VV>10s)} and a \textbf{2.97\% decrease (p< 0.01)} in \textbf{Query Reformulation Rate}, validating the framework's effectiveness in a real-world production environment.
\vspace{-5pt} 
\begin{table}[h]
\centering
\caption{Online A/B testing result.}
\label{tab:ab-testing}
\begin{tabular}{l|cc}
\toprule
Method & VV>10s & Query Reformulation Rate \\
\midrule
WeWrite & +1.07\% & -2.97\% \\
\bottomrule
\end{tabular}
\vspace{-10pt}
\end{table}

\section{Conclusion}
\label{sec:concl}

In this paper, we presented a novel \textbf{Personalized Demand-aware Query Rewriting} framework for video search. 
Addressing the limitations of implicit history modeling, we proposed an explicit generative approach that answers two fundamental questions: \textit{When to Write} and \textit{How to Write}.
Through posterior-based sample mining, we ensure that rewriting is triggered only when necessary. 
By combining SFT with RL, we guarantee that generated queries are both personalized and retrieval-friendly. 
Furthermore, the "Lightweight Recall" architecture overcomes the latency bottleneck of LLMs.
Extensive offline evaluations and online A/B testing confirm that WeWrite effectively resolves user ambiguity, significantly improving search satisfaction and system efficiency.

\begin{acks}
We gratefully acknowledge the Search Application Department, Weixin Group, Tencent, for their insightful discussions and continuous support throughout this project.
\end{acks}

\section*{Speaker Bio}
Chenxing Wang is currently serving as the senior researcher in Weixin Group, Tencent. He received the Ph.D. degree with the School of Computer Science (National Pilot Software Engineering School), Beijing University of Posts and Telecommunications in July, 2025. His current main interests include LLM-based video search and spatio-temporal data mining. He has published several papers in top journals and conference proceedings, such as SIGIR, WWW, ICDE, TMC, T-ITS, IoTJ, IS, IJIS, T-IV and ICANN.

\bibliographystyle{ACM-Reference-Format}
\balance
\bibliography{sample-base}

\end{document}